\documentclass[aip,jcp,preprint]{revtex4-2}
\usepackage{amsmath}
\usepackage{amssymb}
\usepackage{graphicx}
\usepackage[usenames,dvipsnames]{color}

\begin{document}
\title{Numerically ``exact'' simulations of entropy production in the fully quantum regime:\\ Boltzmann entropy versus von Neumann entropy}
\author{Souichi Sakamoto}
\email{sakamoto@kuchem.kyoto-u.ac.jp}
\author{Yoshitaka Tanimura}
\email{tanimura.yoshitaka.5w@kyoto-u.jp}
\affiliation{Department of Chemistry, Graduate School of Science, Kyoto University, Kyoto 606-8502, Japan}
\date{\today}

\begin{abstract}
We present a scheme to evaluate thermodynamic variables for a system coupled to a heat bath under a time-dependent external force using the quasi-static Helmholtz energy from the numerically ``exact'' hierarchical equations of motion (HEOM). We computed the entropy produced by a spin system strongly coupled to a non-Markovian heat bath for various temperatures. We showed that when changes to the external perturbation occurred sufficiently slowly, the system always reached thermal equilibrium. Thus, we calculated the Boltzmann entropy and the von Neumann entropy for an isothermal process, as well as various thermodynamic variables, such as changes of internal energies, heat, and work, for a system in quasi-static equilibrium based on the HEOM. We found that, although the characteristic features of the system entropies in the Boltzmann and von Neumann cases as a function of the system--bath coupling strength are similar, those for the total entropy production are completely different. The total entropy production in the Boltzmann case is always positive, whereas that in the von Neumann case becomes negative if we chose a thermal equilibrium state of the total system (an unfactorized thermal equilibrium state) as the initial state. This is because the total entropy production in the von Neumann case does not properly take into account the contribution of the entropy from the system--bath interaction. Thus, the Boltzmann entropy must be used to investigate entropy production in the fully quantum regime. Finally, we examined the applicability of the Jarzynski equality.
\end{abstract}
\maketitle

\section{INTRODUCTION}
In thermodynamics and statistical mechanics, entropy is an important metric representing the time-irreversible dynamics of an isolated system. The second law of thermodynamics states that the entropy production of an isolated system is always positive, whereas it is zero if the processes are reversible under thermodynamic conditions. Although investigating entropy production in the classical regime is straightforward, for example, using analytical approaches\cite{Gallavotti,BroeckPRE08,JarzynskiAnnu11,SeifertRPP12,SeifertPRL16,Miller} and classical molecular dynamics simulations,\cite{EvansPRL93,CrooksPRE99,Ayton,Mittag} doing so in the quantum regime remains challenging because the microscopic nature of the main system is described by quantized energy states and the dynamics of the system is reversible in time.

In the real world, however, even if the universe has only an isolated excited atom, it will evolve toward thermal equilibrium at 3\,K due to the unavoidable interaction of the system with the surrounding vacuum radiation fields. Thus, a system--bath model, in which a small quantum system is coupled to a bath typically modeled by an infinite number of harmonic oscillators, has been employed to study open quantum dynamics.\cite{Alonso,Ritort,Brandao,Trotzky,Gemmer,Nori1,Brunner,Chotorlishvili,StrasbergPRE19,Thoss} This system--bath model can describe the time irreversibility of the dynamics as the system evolves toward thermal equilibrium in which the energy supplied by fluctuations and the energy lost through dissipation are balanced. The temperature of the bath does not change because its heat capacity is infinite. Moreover, the total energy of the system is conserved if the dynamics described by the total system--bath Hamiltonian is treated properly. To obtain the reduced equations of motion in a compact form, the Markovian assumption is usually employed, in which the correlation time is very short in comparison to the characteristic time of the system dynamics. Widely used approaches for investigating open quantum dynamics employ the Redfield equation and the quantum master equation, which can be derived from the quantum Liouville equation with the full Hamiltonian by reducing the number of degrees of freedom of the heat bath.\cite{Kosloff14,Gelbwaser15,Polkovnikov2011,Korzekwa2016,Hofer2017,Gonzalez2017,Mitchison2017} Several studies on the origin of irreversibility,\cite{Hanggi2005,Harada2005,Saito2008} including a fluctuation theorem, have been developed. \cite{Jarzynski2004,Kurchan,Tasaki,YukawaJPSJ00,CrooksStat08,Campisi09,Campisi11,Mukamel} 

It has been shown, however, that these equations do not satisfy the necessary positivity condition of the population states without imposing a rotating-wave approximation. Because such approximations modify the form of the system--bath interaction, the thermal equilibrium state and the dynamics of the original Hamiltonian are altered.\cite{TanimuraJPSJ06,YTperspective,YTJCP2014,YTJCP2015} Moreover, the majority of previous studies had to adopt a factorized description of the total system, $\hat \rho (t)= \hat  \rho_A (t) \otimes \hat \rho_B^{eq}$, where $ \hat  \rho_A (t)$ is the system density operator and $\hat \rho_B^{eq}$ is the thermal equilibrium state of the bath without the system--bath interaction.\cite{Kosloff14,Gelbwaser15,Polkovnikov2011,Korzekwa2016,Hofer2017,Gonzalez2017,Mitchison2017,EspoPRE06,EspositoNJP10,Sagawa12,Kosloff13} However, this violates the energy conservation of the total system because this assumption ignores the contribution of the energy from the system--bath interaction. This is a fundamental limitation in applying these approaches to investigate entropy production. 

The different definitions of entropy give rise to another complexity. When investigating entropy production, the Boltzmann entropy has been used in the classical case, whereas the von Neumann entropy has been used in the quantum case.\cite{Spohn,Alicki,Yukawa01,Callens,Breuer,EspoPRE06,EspositoNJP10,Sagawa12,Kosloff13} Note that throughout this paper, the Boltzmann entropy refers to the entropy derived from the Helmholtz energy. Although the Boltzmann entropy and the von Neumann entropy coincide when the main system is in thermal equilibrium, they are different when there is an external perturbation that is the source of entropy production.  
The relation between the von Neumann entropy and the second law has been extensively studied, for both thermal equilibrium and nonequilibrium cases.\cite{Alicki, Spohn, Breuer, Yukawa01, Callens, EspoPRE06, EspositoNJP10,Sagawa12, Kosloff13} 
Because the main system is microscopic in the quantum case and because the quantum coherence between the system and bath characterizes the quantum nature of the system--bath dynamics, the role of the system--bath interaction has to be examined carefully. For example, although the factorized thermal equilibrium state, $\hat \rho_{tot}^{eq}= \hat  \rho_A^{eq} \otimes \hat \rho_B^{eq}$, where $\hat  \rho_A^{eq}$ is the equilibrium state of the system without the system--bath interaction, is often employed as an initial state when investigating open quantum dynamics, in real situations, the system and the bath are quantum mechanically entangled (bath entanglement).\cite{YTperspective} As we will illustrate below, the contribution of the entropy from the system--bath interaction must be taken into account for the total entropy production, otherwise the second law of thermodynamics is violated, in particular if there is strong system--bath coupling at low temperatures.

In the present paper, we examine the role of the system--bath interaction for entropy production and various thermodynamic variables by computing the von Neumann entropy and the Boltzmann entropy. For this purpose, we employ the real-time hierarchical equations of motion (HEOM).\cite{KATO2015,KATO2016,Katobook2019,TanimuraJPSJ06,YTperspective,YTJCP2014,YTJCP2015,Tanimura89A,TanimuraPRA90,IshizakiJPSJ05,TanimuraPRA91,TanimuraJCP92, KatoJPCB13} Because the HEOM can provide an ``exact'' numerical treatment of the dynamics defined by a system--bath Hamiltonian, it is possible to carry out desktop experiments to verify quantitatively fundamental aspects of the quantum thermodynamics. The effects of non-perturbative and entangled system--bath interactions have been investigated with the HEOM approach, based on quantum thermodynamics, for heat currents\cite{NitzanHC, CaoJPCC19,CaoJPCL20,KATO2015,KATO2016,Katobook2019} and heat engines.\cite{NitzanHM,Nori0,Nori,Aspuru,Newman,Segal19}
Here, we investigate the effects of the system--bath interaction on entropy production using a spin-boson model for various values of the system--bath coupling using a factorized and a true (unfactorized) thermal equilibrium state as the initial state. Although the HEOM have been used to investigate the production of von Neumann entropy,\cite{Kawai2019} here we study not only the von Neumann entropy but also the Boltzmann entropy, as well as various thermodynamic variables, including the changes of internal energies, heat, and work, by simulating the kinetic response of the system. We then show that entropy production in the von Neumann case becomes negative if we calculate it for a correlated thermal equilibrium state because no entropy is contributed by the system--bath interaction. In contrast, entropy production in the Boltzmann case is always positive. 

We should emphasize that, although our scheme to calculate thermodynamic variables is based on a desktop experiment to measure the kinetic response of a system using the real-time HEOM, a similar scheme could be used in a real experiment to determine the thermodynamic variables of a quantum system in a thermal environment, by applying a slowly changing external field. In this regard, our approach is like the theory developed by Jarzynski.\cite{Jarzynski2004}

This paper is organized as follows. In Sec.~\ref{sec:model}, we introduce the system--bath Hamiltonian and the HEOM formalism. We then describe the scheme used to calculate various thermodynamic variables based on open quantum dynamics theory. In Sec.~\ref{sec:results}, we present the numerical results of the system entropy and entropy production as a function of the system--bath coupling strength for the Boltzmann and von Neumann cases. To analyze the difference between these two cases, we calculate various thermodynamic variables. Section~\ref{sec:conclusion} contains concluding remarks.

\section{System--bath model and thermodynamic variables}
\label{sec:model}

\subsection{Hamiltonian}
\label{subsec:Hamiltonian}

To carry out desktop experiments to verify fundamental aspects of quantum thermodynamics in a practical manner, we consider a system $A$ coupled to a heat bath $B$ of harmonic oscillators. The Hamiltonian of the total system is given by:
\begin{align}
    {\mathcal {\hat H}}(t) = {\mathcal {\hat H}}_A(t) + {\mathcal {\hat H}}_I + {\mathcal {\hat H}}_B,
    \label{eq:Hamiltonian}
\end{align}
where $ {\mathcal {\hat H}}_A(t) = {\mathcal {\hat H}}_A^0 +  {\mathcal {\hat H}}_E (t)$ with $ {\mathcal {\hat H}}_E (t)=0$ for $t \le 0$. Here, ${\mathcal {\hat H}}_A^0$ is the system Hamiltonian, whose explicit time dependence originates from the coupling with the external driving field, $ {\mathcal {\hat H}}_E (t)$. 
The bath Hamiltonian ${\mathcal {\hat H}}_B$ can be expressed as:
\begin{align}
    {\mathcal {\hat H}}_B = \sum_j \left[ \frac{\hat p_j^2}{2 m_j} + \frac{1}{2} m_j \omega_j^2 {\hat x_j}^2 \right],
    \label{eq:bathHamiltonian}
\end{align}
where $\hat p_j$, $\hat x_j$, $m_j$, and $\omega_j$, are the momentum, position, mass, and frequency of the $j$th bath oscillator, respectively, and 
the system--bath interaction ${\mathcal {\hat H}}_I$ is given by ${\mathcal {\hat H}}_I ={\hat V} \sum_j g_j {\hat x}_j$, where $\hat V$ is the system part of the interaction and $g_j$ is the coupling constant between the system and the $j$th bath oscillator. 
The heat bath can be characterized by the spectral distribution function (SDF), defined by
\begin{align}
  J (\omega) \equiv \sum_{j=1}^{N}\frac{\pi g_{j}^2}{2m_{j} \omega_{j} } \delta(\omega-\omega_{j}),
  \label{eq:J_wgeneral}
\end{align}
and the inverse temperature $\beta \equiv 1/k_{\mathrm{B}}T$, where $k_\mathrm{B}$ is Boltzmann's constant. Various environments, for example, those consisting of nanostructured materials, solvents, or protein molecules, can be modeled by adjusting the form of the SDF.\cite{YTperspective}
For the heat bath to be an unlimited heat source with an infinite heat capacity, the number of heat bath oscillators $N$ can be made infinitely large by replacing $J (\omega)$ with a continuous distribution. In the present treatment, although the number of degrees of freedom for the bath is infinite, the total energy described by Eq.~\eqref{eq:Hamiltonian}, including the work done by the external force, is conserved when we rigorously treat not only the system but also the bath. The full system--bath model can be regarded as an isolated system.

\subsection{Reduced density matrix and the hierarchical equations of motion}

The reduced density matrix is defined by 
\begin{align}
\hat \rho_A (t) = {\operatorname{tr}_{B}} \left\{ \exp_{+} \left[-\frac{i}{\hbar} \int_0^t dt{\mathcal {\hat H}}(t) \right] \hat \rho_{A+B}^{eq} \exp_{-} \left[\frac{i}{\hbar} \int_0^t dt{\mathcal {\hat H}}(t) \right] \right\},
 \label{eq:rho_S}
\end{align}
where $\exp_{-}$ and $\exp_{+}$ are the time-ordered exponentials, and ${\hat \rho_{A+B}^{eq}}$ is the thermal equilibrium state of the system. As an initial condition, here we consider the factorized thermal equilibrium state and the correlated (true) thermal equilibrium state expressed as ${\hat {\rho'}_{A+B}^{eq}} = e^{-\beta{\mathcal {\hat H}}_A^0} \otimes e^{-\beta {\mathcal {\hat H}_B}} / Z_A^0 Z_B$ and ${\hat \rho_{A+B}^{eq}} \equiv e^{-\beta({\mathcal {\hat H}}_A^0 + {\mathcal {\hat H}_I + {\mathcal {\hat H}_B}})} / Z_{tot}^0$, respectively, where $Z_A^0 \equiv {\operatorname{tr}_A} \{ e^{-\beta {\mathcal {\hat H}}_A^0} \}$, $Z_B \equiv {\operatorname{tr}_B} \{ e^{-\beta {\mathcal {\hat H}_B}} \}$, and $Z_{tot}^0 = Z_{tot} (\tau=0)$ with $Z_{tot} (\tau) \equiv {\operatorname{tr}_{A+B}} \{ e^{-\beta({\mathcal {\hat H}_A}(\tau) + \mathcal {\hat H}_I + {\mathcal {\hat H}_B})} \}$.
In the path integral representation, $\hat \rho_A (t)$ can be evaluated from the initial conditions described by the correlated thermal equilibrium state\cite{YTJCP2014,Grabert} and the factorized thermal state.\cite{Feynman63} 

The effects of the bath on the system are characterized by the noise correlation function $C(t) = \langle {\hat X}(t) {\hat X}(0) \rangle_B$, where the operator ${\hat X}$ is the collective bath coordinate 
defined by ${\hat X} = \sum_j g_j {\hat x}_j$. Here, the notation $\langle \ldots \rangle_B$ represents the average taken with the canonical distribution of the bath. 
The noise correlation function is expressed through $J(\omega)$ as
\begin{align}
    C(t) = \hbar \int_0^{\infty} \frac{d \omega}{\pi} J(\omega) \left[ \coth \left( \frac{1}{2}\beta \hbar \omega \right) \cos(\omega t) - i \sin(\omega t) \right].
    \label{eq:correlation}
\end{align}
We assume that the SDF is given by the Drude distribution, $J (\omega) = \eta \gamma^2 \omega / (\omega^2 + \gamma^2)$, where $\eta$ is the system--bath coupling strength and $\gamma$ is the cutoff frequency. 
Then, $C(t)$ is expressed in terms of exponential functions and a delta function as $C(t) = \sum_{k=0}^{L} (c'_k + ic''_k) \gamma_k e^{- \gamma_k t} + 2\Delta_L \delta (t)$, where $c'_k$, $c''_k$, $\gamma_k$, and $\Delta_L$ are constants. 
This form of $C(t)$ allows us to derive the HEOM, which consists of the following sets of equations of motion with the auxiliary density operators (ADOs):
\cite{Tanimura89A,TanimuraPRA90,IshizakiJPSJ05,TanimuraJPSJ06,YTperspective,YTJCP2014,YTJCP2015}
\begin{multline}
    \frac{\partial}{\partial t} {\hat \rho}_{(n_0,\ldots,n_L)}(t) = -\left[ \frac{i}{\hbar} {\mathcal {\hat L}}_A(t) + \Delta_L {\hat \Phi}^2 + \sum_{k=0}^{L} n_k \gamma_k\right] {\hat \rho}_{(n_0,\ldots,n_L)}(t) \\
     +\sum_{k=0}^{L} n_k {\hat \Theta_k} {\hat \rho}_{(\ldots,n_k - e_k,\ldots)}(t) + {\hat \Phi_k} \sum_{k=0}^{L} {\hat \rho}_{(\ldots,n_k + e_k,\ldots)}(t).
    \label{eq:HEOM}
\end{multline}
Here, $\mathbf{e}_k$ is the unit vector along the $k$th direction, and we have defined the superoperators ${\mathcal {\hat L}}_A(t) {\mathcal {\hat O}} \equiv \left[ {\mathcal {\hat H}_A}(t), {\mathcal {\hat O}} \right]$, $\hat \Phi {\mathcal {\hat O}} \equiv {i} \left[ {\hat V}, {\mathcal {\hat O}} \right]/ {\hbar}$, and $\hat \Theta \equiv c_k' {\hat \Phi} - c_k'' {\hat \Psi}$ with $\hat \Psi {\mathcal {\hat O}} \equiv  \{ {\hat V}, {\mathcal {\hat O}} \}/ {\hbar}$ for any operator ${\mathcal {\hat O}}$.
Each ADO is specified by the index ${\bf n} = (n_0, \ldots,n_L)$, where each element takes a non-negative integer value.
The ADO for ${\bf n} = {\bf 0}$ corresponds to the actual reduced density operator. 
In the HEOM approach, the factorized initial state is set as ${\hat \rho}_{{\bf n} = {\bf 0 }}(0)= \exp[ -\beta {\mathcal {\hat H}}_A^0 ]/Z_{A}^0$ and all the other hierarchical elements are set to zero. The correlated thermal equilibrium state can be set by running the HEOM program for fixed ${\mathcal {\hat H}}_A (t)= {\mathcal {\hat H}}_A^0$ until all of the hierarchical elements reach a steady state ${\hat \rho}_{\bf n}^{eq}={\hat \rho}_{{\bf n}}(t\rightarrow \infty)$. Then, we use these elements as the initial state ${\hat \rho}_{\bf n}(0)={\hat \rho}_{\bf n}^{eq}$. The steady-state solution of the first hierarchical  element agrees with the correlated thermal equilibrium state defined by $\hat{\rho}_A^{eq}= {\operatorname{tr}_B} \{\exp(-\beta {\mathcal {\hat H}} (0))\}/Z_{tot}^0$, whereas the other elements describe bathentanglement states.\cite{YTperspective,YTJCP2014,YTJCP2015}
We compute various thermodynamic variables as the change of the equilibrium distributions by numerically integrating the HEOM with respect to time using the fourth-order low-storage Runge--Kutta method.\cite{LSRK42017,Ikeda2018CI}

\subsection{Quasi-static Helmholtz energy and Boltzmann entropy}
 \label{subsec:Z}

We consider the partition function of the reduced system at the fixed snapshot time $\tau$, defined as 
$Z_A (\tau) \equiv {\rm tr_{A+B}} \{ e^{-\beta {\mathcal {\hat H}}(\tau)} / Z_B^{\tau} \}$, where $Z_B^{\tau}$ is the bath part of the partition function which is reduced from the total Hamiltonian. In practice, however, we may set $Z_B^{\tau}= Z_B$, because we only need the ratio $Z_A (\tau)/ Z_A (0)$ that is evaluated from the real-time HEOM, as we will show below.
In the functional integral form, this is expressed as:
\cite{Leggett1981,Leggett1987,YTJCP2014,YTJCP2015}
\begin{align}
  Z_{A}(\tau) \equiv  \int d{\bf \sigma} \int_{{\bf \sigma}(0)= {\bf \sigma}}^{{\bf \sigma}(\beta\hbar)= {\bf \sigma}} D[{\bf \sigma}(u)] \exp\left[ -\frac{1}{\hbar} S_A[{\bf \sigma}(u); \tau] \right],
  \label{eq:ImDensity}
\end{align}
where ${\bf \sigma}(u)$ is the functional form of the spin operators, which are described using Grassmann variables at the inverse temperature $u$ and 
\begin{equation}
S_A[{\bf \sigma}(u); \tau] = \int_{0}^{\beta\hbar} d u'H_A(u';\tau) \\ 
   - \frac1{\hbar} \int_{0}^{\beta\hbar} d u'' \int_{0}^{u''} d u'  { V} (u'') { V} (u') \bar L (u'' - u' ),
\label{eq:PhitA2}
\end{equation}
and 
\begin{align}
\bar L(u') = \hbar \int_0^{+\infty} \frac{d \omega}{\pi} J(\omega)
 \frac{\cosh\left( \beta \hbar\omega/2 - \omega u' \right)}
{\sinh \left( \beta \hbar\omega/2 \right)}.
\label{eq:P1}
\end{align}
Here, $H_A(u;\tau)$ and $V(u)$ are the functional representations of $\mathcal {\hat H}_A(\tau)$ and ${\hat V}$. 

As we will demonstrate numerically below, when $\mathcal{\hat H}_E (t)$ changes much more slowly than the relaxation time of the system, the reduced density operator $\hat \rho_A (t)$, evaluated with Eq.~\eqref{eq:HEOM}, approaches the quasi-thermal equilibrium state of the system at time $t=\tau$ as ${\hat \rho}_{A}^{qeq} (\tau) \approx {\operatorname{tr}_{B}} \{ e^{-\beta({\mathcal {\hat H}_A}(\tau) + \mathcal {\hat H}_I + {\mathcal {\hat H}_B})} \}/ Z_{tot} (\tau)$.

Although $Z_{A}(\tau)$ can be evaluated from the imaginary-time HEOM by calculating the system partition function,\cite{YTJCP2014,YTJCP2015} the numerical integration is not easy, especially for lower temperatures, due to the oscillatory nature of the noise correlation function in imaginary time. Moreover, to calculate the change of the thermodynamic variables, including entropy production by an isothermal process, we need only the ratio $Z_{A} (t) / Z_{A} (0)$. Thus, we use the quasi-static solution ${\hat \rho}_{A}^{qeq}(t)$, calculated from the real-time HEOM, to evaluate the derivative of $\ln( Z_{A} (t))$ with respect to time $t$:
\begin{align}
    \frac{\partial}{\partial t} \left( -\frac{1}{\beta} \ln Z_{A}(t) \right) = {\operatorname{tr}_{A}} \left\{ {\hat \rho}_A^{qeq} (t) \frac{\partial}{\partial t} {\mathcal {\hat H}_A}(t) \right\},
    \label{eq:qspower}
\end{align}
where the right-hand side (RHS) of the above equation corresponds to the power in the quasi-static isothermal process. A derivation of Eq.~(\ref{eq:qspower}) is presented in Appendix~\ref{sec:qspower}. From the definition of the Helmholtz energy, $F \equiv - \ln Z/\beta$, the change of the ``quasi-static'' Helmholtz energy at time $\tau$ is expressed as:
\begin{align}
    \Delta F_A(\tau) \equiv \int_0^{\tau} {\operatorname{tr}_A} \left\{ {\hat \rho}_A^{qeq} (t) \frac{\partial}{\partial t} {\mathcal {\hat H}_A}(t) \right\} dt.
    \label{eq:Fchange}
\end{align}
Here, the RHS of the above equation is the quasi-static work done on the system during the isothermal operation, which agrees with the work for the quasi-static equilibrium process as $\Delta F_A(\tau)=W^{qeq} (\tau) $. From the above, the change of the ``quasi-static'' Boltzmann entropy $\Delta S_A(\tau)$ is obtained as:
\begin{align}
    \Delta S_A(\tau) = k_B \beta^2 \frac{\partial}{\partial \beta} \Delta F_A(\tau).
    \label{eq:Bchange}
\end{align}
Note that this definition of the system entropy includes a contribution from the system part of the system--bath interaction. 
Accordingly, the change of the internal energy is evaluated~as:
\begin{align}
  \Delta U_A(\tau) =\frac{\partial}{\partial \beta} \left(\beta \Delta F_A(\tau)\right).
 \label{eq:UA}
\end{align}
The work $W(\tau)$ is expressed as
\begin{align}
    W(\tau) = \int_0^{\tau} P(t) dt,
     \label{eq:work}
\end{align}
with the power defined as:
\begin{align}
    P(t) \equiv {\operatorname{tr}_A} \left\{ {\hat \rho}_{A} (t) \frac{\partial}{\partial t} {\mathcal {\hat H}_A}(t) \right\}.
    \label{eq:power}
\end{align}
The work described by Eq.~\eqref{eq:work} is equivalent to the change of the total system energy during the isothermal operation from $t=0$ to $t=\tau$ because the power can also be expressed as $P(t)=\partial U_{tot}(t) /\partial t$, where the total energy is defined as:\cite{KATO2016,Katobook2019,Allahverdyan}
\begin{align}
     U_{tot}(t)\equiv {\operatorname{tr}_{A+B}} \left\{ {\hat \rho}_{tot} (t) \left({\mathcal {\hat H}_A}(t) + {\mathcal {\hat H}_I} + {\mathcal {\hat H}_B}\right) \right\}. 
    \label{eq:energyconservation}
\end{align}
For the system described above, the first law of thermodynamics states that:
\begin{align}
    \Delta Q(\tau) = \Delta U_A(\tau) - W(\tau),
     \label{eq:first}
\end{align}
where $\Delta Q(\tau)$ is the heat released from the bath. 
The total entropy production is then expressed as:
\begin{align}
    \Sigma_{tot}^B(\tau) = k_B^{-1} \Delta S_A(\tau) - \beta \Delta Q(\tau).
    \label{eq:S-production}
\end{align}

To analyze $\Delta Q(\tau)$ more precisely, we further introduce the change of the bath energy (the bath--heat current) at time $\tau$ expressed as $\Delta \langle {\mathcal {\hat H}_B}(\tau) \rangle $. 
In the HEOM formalism, this is evaluated from the first-order hierarchical elements:\cite{KATO2016, Katobook2019}
\begin{align}
\Delta \langle {\mathcal {\hat H}_B}(\tau) \rangle \equiv \int_0^{\tau} \frac{d}{dt} \langle {\hat {\mathcal H}_B}(t) \rangle dt,
\label{eq:bathE}
\end{align}
where
\begin{multline}
    \frac{d}{dt} \langle {\hat {\mathcal H}_B}(t) \rangle = -\frac{2}{\hbar} \operatorname{Im}[C(0)] {\operatorname{tr}_A} \{ {\hat V}^2 {\hat \rho}_A (t) \} - \left( \frac{i}{\hbar} \right)^2 \Delta_L {\operatorname{tr}_A} \{ [[{\hat {\mathcal H}_A}(t),{\hat V}],{\hat V}] {\hat \rho}_A (t) \} \\
  - \sum_{k=0}^{L} \gamma_k {\operatorname{tr}_A} \{ {\hat V} {\hat \rho}_{\mathbf{e}_k} (t) \}.
  \label{eq:BHC}
\end{multline}
Accordingly, the interaction energy
$\langle {\hat {\mathcal H}_I}(t) \rangle \equiv {\operatorname{tr}}_{A+B} \{ {\hat {\mathcal H}_I} {\hat \rho}_{tot} (t) \}$ is evaluated in the HEOM formalism as:\cite{YTJCP2014,KATO2016,Katobook2019}
\begin{align}
    \langle {\hat {\mathcal H}_I}(t) \rangle = - \sum_{k=0}^{L} {\operatorname{tr}_A} \{ {\hat V} {\hat \rho}_{\mathbf{e}_k} (t) \}.
    \label{eq:InteractionE}
\end{align}
Then the change of the interaction energy is evaluated as: $\Delta \langle {\mathcal {\hat H}_I}(\tau) \rangle =\langle {\hat {\mathcal H}_I}(\tau) \rangle - \langle {\hat {\mathcal H}_I}(0) \rangle $. 
The change of the system energy without the system--bath interaction is given by: $\Delta \langle {\mathcal {\hat H}_A}(\tau) \rangle = {\operatorname{tr}_A} \{ {\mathcal {\hat H}_A}(\tau) {\hat \rho}_A^{qeq}(\tau) -{\mathcal {\hat H}_A}(0) {\hat \rho}_A^{qeq}(0)\}$. 
Using the above results with Eqs.~\eqref{eq:work}--\eqref{eq:energyconservation}, we can also evaluate the work from the HEOM:
\begin{align}
W(\tau)= \Delta \langle {\hat {\mathcal H}_A}(\tau) \rangle+ \Delta \langle {\hat {\mathcal H}_I}(\tau) \rangle + \Delta \langle {\hat {\mathcal H}_B} (\tau)\rangle.
    \label{eq:Work_Htot}
\end{align}
Thus, the total entropy production in the Boltzmann case, as presented in Eq.~\eqref{eq:S-production}, can be rewritten as:
\begin{align}
    \Sigma_{tot}^B(\tau) = k_B^{-1} \Delta S_A(\tau) +\beta \Delta \langle {\hat {\mathcal H}_B}(\tau)\rangle+\beta\left( \Delta \langle {\hat {\mathcal H}_I}(\tau) \rangle    - \delta U_A'(\tau) \right),
    \label{eq:S-production2}
\end{align}
where $\delta U_A'(\tau) \equiv \Delta U_A(\tau) - \Delta \langle {\hat {\mathcal H}_A}(\tau) \rangle$ represents the energy of the system part of the system--bath interaction. 

For very weak system--bath interactions, $\Delta \langle {\hat {\mathcal H}_I}(\tau) \rangle$ can be ignored and $\delta U_A'(\tau)$ approaches zero. 
This assumption is often employed in quantum thermodynamics. In reality, however, a system cannot reach thermal equilibrium state on its own without the system--bath interaction because the microscopic nature of the main system is described by quantized states and the dynamics of the system itself is reversible in time. Thus, a careful treatment of $\Delta \langle {\hat {\mathcal H}_I}(\tau) \rangle$ and $\delta U_A'(\tau)$ is necessary.

\subsection{von Neumann entropy}

The von Neumann entropy is commonly used in quantum thermodynamics. It is defined~as:
\begin{align}
S_A^{vN}(t) = -{\operatorname{tr}_A} \{ {\hat \rho_A (t)}\ln{\hat \rho_A (t)} \},
 \label{eq:Neumann-entropyt}
\end{align}
where ${\hat \rho_A (t)}$ is the reduced density matrix. Then the change of the system entropy is given by $\Delta S_{A}^{vN}(\tau) = S_{A}^{vN} (\tau) - S_{A}^{vN} (0)$.  Note that this entropy is consistent with the Boltzmann entropy in thermal equilibrium when the system--bath interaction is very weak. For the von Neumann entropy, entropy production is defined as:
\begin{align}
    \Sigma_{tot}^{vN} (\tau) = \Delta S_{A,qeq}^{vN} (\tau) + \beta \Delta \langle {\mathcal {\hat H}_B} (\tau) \rangle,
    \label{eq:Neumann-production}
\end{align}
where the second term on the RHS is the contribution of the entropy from the bath and where
\begin{align}
\Delta S_{A,qeq}^{vN}(\tau) = - {\operatorname{tr}_A} \{ {\hat \rho_A^{qeq}}(\tau)\ln{\hat \rho_A^{qeq}}(\tau) \}  + {\operatorname{tr}_A} \{ {\hat \rho_A^{qeq}}(0)\ln{\hat \rho_A^{qeq}}(0) \}  
    \label{eq:Neumann-entropy}
\end{align}
is the change of the von Neumann entropy of a system in a quasi-static equilibrium state. Note that, although the contribution of the entropy from the bath in the von Neumann case is defined as being from the bath itself, in the Boltzmann case, it includes the contribution from the system--bath interaction described by the third term of the RHS of Eq.~\eqref{eq:S-production2}. Although the definition of Eq.~\eqref{eq:Neumann-entropy} has been extensively used under various conditions,\cite{Spohn,Alicki,Yukawa01,Callens,Breuer,EspoPRE06,EspositoNJP10,Sagawa12,Kosloff13}  as we will show in the next section, the positivity of entropy production in the von Neumann case breaks due to the contribution from the system--bath interaction, if the initial equilibrium state is correlated.

\section{NUMERICAL RESULTS}
 \label{sec:results}

\subsection{Real-time responses}

Our scheme for calculating thermodynamic variables, which we described in Sec.~\ref{sec:model}, is based on a simulation of the kinetic response under an external perturbation using the HEOM. Although the HEOM are applicable for a range of systems,\cite{YTperspective} including chemical reactions,\cite{TanimuraPRA91,TanimuraJCP92} quantum ratchets,\cite{KatoJPCB13} spin glass,\cite{Tsuchimoto2015,Nakamura2018} and photosynthesis,\cite{Schuten11,KramerFMO2DLorentz,Fujihashi2015,Sakamoto2017} here we employ a simple spin-boson system for demonstration. For a system of this kind, special attention has to be paid to the role of the system--bath interaction, because the main system cannot reach thermal equilibrium by itself without the system--bath interaction, even if the interaction is very weak. Here, we set the system Hamiltonian:
\[
{\mathcal {\hat H}}_A^0 = \frac{1}{2}\hbar \omega_0 (|e\rangle \langle e|-|g\rangle \langle g|)
\]
and ${\mathcal {\hat H}_E}(t) = f(t) \hbar \omega_0 (|e\rangle \langle e|-|g\rangle \langle g|)$, where $|e\rangle$ and $|g\rangle$ are the excited state and ground state, respectively.
The system--bath interaction is defined as $\hat V =  |g\rangle \langle e|+|e\rangle \langle g|$. In the simulation, we set $\eta=1$ and $\beta \hbar \omega_0 = 1$. 
Throughout this paper, we fix the cutoff frequency $\gamma=\omega_0$, which corresponds to a moderate non-Markovian case. 
Note that even if $\gamma$ is very large, the noise is non-Markovian in the very low temperature regime due to quantum thermal fluctuations, as demonstrated for the simulation of muon spin spectroscopy ($\mu$SR).\cite{TakahashiJPSJ20}
The system is driven by the external field:
\begin{align}
    f(t)=
    \begin{cases}
    0, & t \leq 0, \\
    \dfrac{1}{4T} t, & 0 < t \leq T, \\
    \dfrac{1}{4}, & T < t, \\
    \end{cases}
  \label{eq:f(t)}
\end{align}
where $T$ is the time duration of the driving force. 
We first investigate the response of the thermodynamic variables under an external field with different growth rates over time. We evaluated the von Neumann entropy at time $\tau$ from Eq.~\eqref{eq:Neumann-production} by numerically integrating the HEOM until time $t=\tau$, starting from the correlated equilibrium state at $t=0$ and then using the zeroth element of the solution, ${\hat \rho}_{\mathbf{n}=0} (\tau)$. 

\begin{figure}[tp]
    \centering
     \includegraphics[width = 8.0cm]{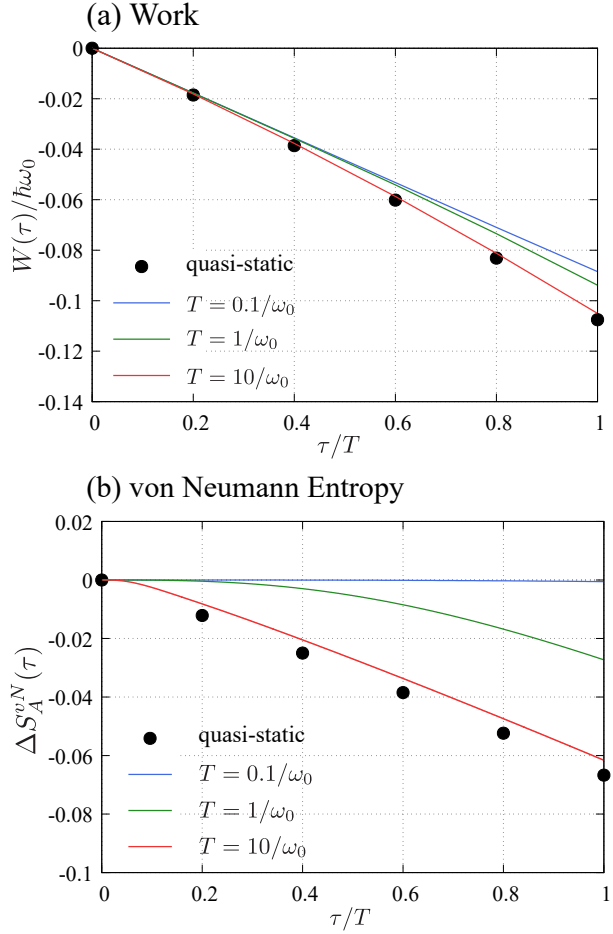}
     \caption{(a) The work $W(\tau)$ calculated from Eq.~(\ref{eq:work}) and (b) the change of the von Neumann entropy of the main system $\Delta S_A^{vN}(\tau)$ calculated from Eq.~\eqref{eq:Neumann-entropyt}. Both are plotted as functions of time $\tau/T$ for fixed $\eta=1$ and $\beta \hbar \omega_0 = 1$. The black dots represent the results from the quasi-static distribution ${\hat \rho_{A}^{qeq}}(\tau)$ that satisfy $W^{qeq}(\tau)=\Delta F_A (\tau)$. The colored curves represent different time durations: $T=0.1/\omega_0$ (blue curve), $T=1/\omega_0$ (green curve), and $T=10/\omega_0$ (red curve).}
     \label{fig:TD}
\end{figure}

In Fig.~\ref{fig:TD}, we depict the results calculated for (a) the work $W(\tau)$ and (b) the change of the von Neumann entropy $\Delta S_A^{vN}(\tau)$ at time $\tau$. The black dots are the values calculated from the quasi-static distribution ${\hat \rho_{A}^{qeq}}(\tau)$. 
First, note that the von Neumann entropy becomes large when the ground state and excited states are equally populated, 
whereas it becomes small if the population is localized in the ground state. In the present case, 
because the external field enhances the excitation energy, the population of the excited state decreases as $\tau$ increases. Because $\Delta S_A^{vN}(\tau)$ is zero at $\tau=0$, the change in the entropy is negative. 

When the external perturbation is applied very slowly, the work and the changes of the von Neumann entropy approach their quasi-static equilibrium values, which are represented by the black dots. Note that if we increase the system--bath coupling $\eta$, instead of $T$, the von Neumann entropy is suppressed to the quasi-static value more rapidly (not shown). These results imply that the reduced density operator, defined in Eq.~\eqref{eq:rho_S}, coincides with ${\hat \rho_{A}^{qeq}} (\tau)$ at each time $\tau$ if the perturbations are sufficiently slow or if the system--bath coupling is strong. A slower external perturbation or stronger system--bath coupling results in smaller work $W(\tau)$.  The lower limit of the work is identical to the change of the quasi-static Helmholtz energy, i.e., $W^{qeq}(\tau)=\Delta F_A (\tau)$.  We, thus, have $\Delta F_A (\tau) \leq W (\tau)$, which corresponds to the second law of thermodynamics. 

\subsection{System entropy and entropy production calculated from the correlated equilibrium state}
\label{subsec:EntropyProduction}

Based on the above results, here we calculate both the Boltzmann entropy and the von Neumann entropy using ${\hat \rho_{A}^{qeq}}(\tau)$, which was obtained as the steady-state solution of the HEOM by integrating them from $t=0$ to sufficiently long time $t\gg 1/\omega_0$ for ${\mathcal {\hat H}}(\tau)$ with the fixed time $\tau$. 
We calculated the change of the quasi-static Helmholtz energy $\Delta F_A(\tau)$ from Eq.~\eqref{eq:Fchange}. Then the change of the system entropy $\Delta S_A$ was calculated from Eq.~(\ref{eq:Bchange}) by numerically differentiating $\Delta F_A(\tau)$ with respect to $\tau$ using a seven-point finite difference method with grid spacing $\Delta \beta = 0.01 / \hbar \omega_0$. Using the first law of thermodynamics Eq.~\eqref{eq:first} with Eqs.~\eqref{eq:UA} and~\eqref{eq:work}, we evaluated the total entropy production $\Sigma_{tot}^B$ from  Eq.~(\ref{eq:S-production}). In the von Neumann case, the change of the system entropy $\Delta S_A^{vN}$ and total entropy production $\Sigma_{tot}^{vN}$ were calculated from Eqs.~(\ref{eq:Neumann-entropy}) and~(\ref{eq:Neumann-production}), respectively. The quasi-static Helmholtz energy, as well as various thermodynamic variables, were obtained by numerically simulating the time evolution of the system from the true (correlated) thermal equilibrium state, ${\hat \rho_A} (0)={\operatorname{tr}_B} \{ e^{-\beta({\mathcal {\hat H}}_A^0+ \mathcal {\hat H}_I + {\mathcal {\hat H}_B})} \} / Z_{tot}^0$, to the final state ${\hat \rho_A} (+\infty) = {\hat \rho_A^{qeq}} (T)$. Here, we consider the slow perturbation case, $T=10/\omega_0$.

\begin{figure}[tb]
\centering
\includegraphics[width = 1.0\textwidth]{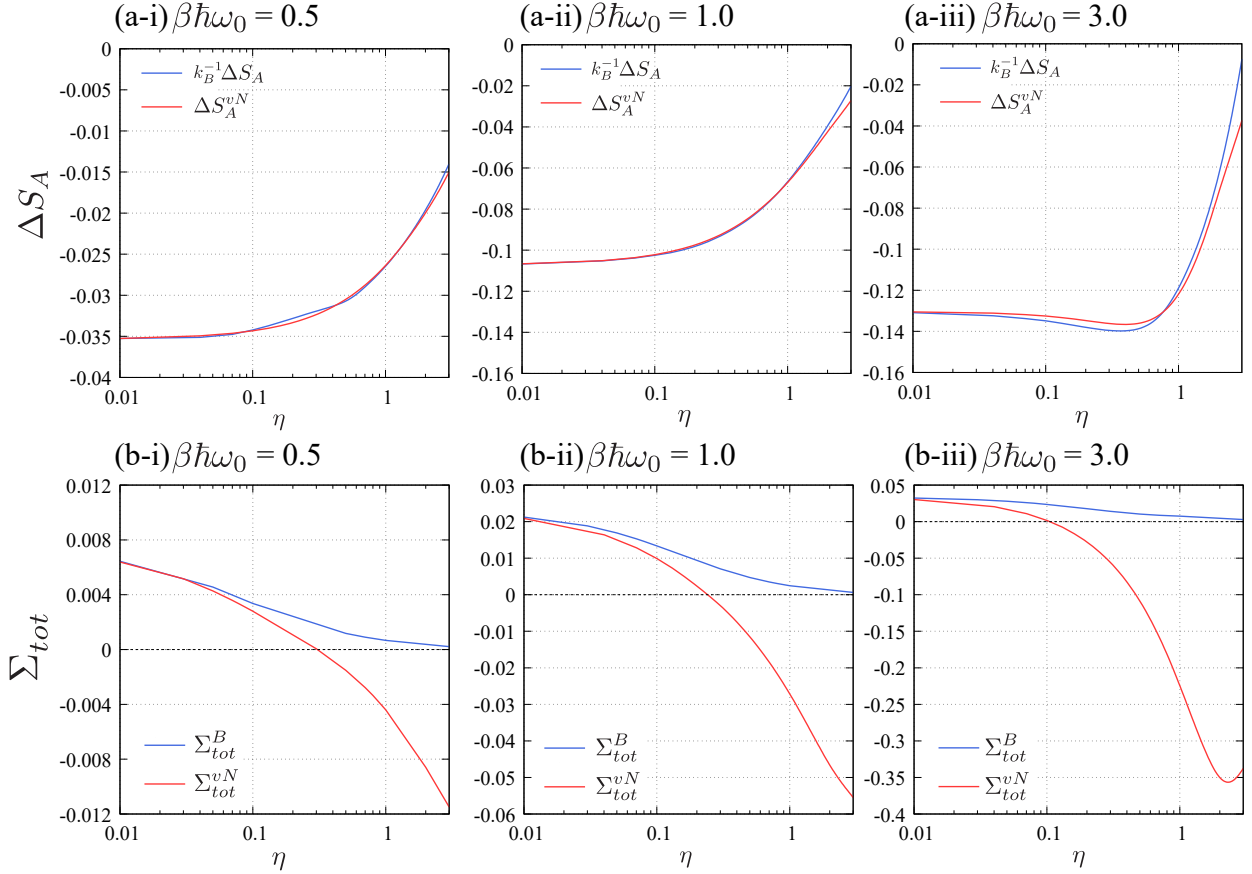}
\caption{(a) Change of the system entropy and (b) the total entropy production in the Boltzmann case (red curves) and the von Neumann case (blue curves) as functions of the system--bath coupling strength $\eta$, for the inverse temperatures: (i) $\beta \hbar \omega_0 = 0.5$, (ii) $\beta \hbar \omega_0 = 1$, and (iii) $\beta \hbar \omega_0 = 3$.} 
      \label{fig:entropy}
\end{figure}

In Fig.~\ref{fig:entropy}, we illustrate: (a) the change of the system entropy $\Delta S_A$ and (b) the total entropy production $\Sigma_{tot}$ in the Boltzmann (blue) and von Neumann (red) cases, as functions of the system--bath coupling strength $\eta$ for different temperatures. We first discuss $\Delta S_A$, as shown in Figs.~\ref{fig:entropy}(a-i)--\ref{fig:entropy}(a-iii). As explained for Fig.~\ref{fig:TD}, $\Delta S_A$ becomes negative because the external force enhances the excitation energy and thus, the population is localized in the ground state. This tendency becomes prominent at lower temperatures due to the small thermalization that arises from the fluctuations of the bath. The differences between the Boltzmann and von Neumann cases become larger as the system--bath coupling strengthens, in particular at low temperatures (see also Ref. \onlinecite{YJYang20}), but the overall profiles for the two results are similar. This is because we calculated both entropies using the reduced density matrix of the main system obtained from the HEOM, and thus, the effects of the non-perturbative system--bath interaction were indirectly taken into account in the von Neumann case.

In all cases in Figs.~\ref{fig:entropy}(a-i)--\ref{fig:entropy}(a-iii), the changes of the system entropy increase as the system--bath coupling strength increases because the excited state is populated in the strong system--bath coupling region. At low temperatures [Fig.~\ref{fig:entropy}(a-iii)], however, the increase in the entropy change is suppressed in the region $0.05 \le \eta \le 0.5$. This is due to the relaxation of the excited state arising from the dissipation, whereas the thermal excitation arising from the fluctuations is suppressed in this low-temperature regime. Thus, the ground state population is more localized and, as a result, the entropy becomes small in this parameter region. For very large $\eta$, however, the system and the bath are strongly coupled and the energy eigenstates of the system become continuous. Thus, the change of the system entropy becomes large for large $\eta$.

Although the changes of the system entropy in the Boltzmann and von Neumann cases are mostly determined by the ground- and excited-state populations and are not sensitive to the definition of the entropy, the total entropy production in these two cases exhibits completely different behavior, as illustrated in Figs.~\ref{fig:entropy}(b-i)--\ref{fig:entropy}(b-iii).
The total entropy production in the Boltzmann case is always positive, whereas that in the von Neumann case becomes negative, even in a weak coupling region. 
This difference is due to the third term on the RHS of Eq.~\eqref{eq:S-production2}, $\beta( \Delta \langle {\hat {\mathcal H}_I} \rangle - \delta U_A')$, in the Boltzmann expression. 
Note that, in quantum information theory, the difference of the von Neumann entropy is the quantum mutual information. It is defined by $I(A:B) \equiv S_A^{vN} + S_B^{vN} - S_{tot}^{vN}$ and is employed as a measure of the correlation between the quantum states of the system and the bath.\cite{Goold} 
In the present case, although we cannot evaluate the bath part of the von Neumann entropy directly, from the difference between Eqs.~\eqref{eq:S-production2} and~\eqref{eq:Neumann-production}, it should be reasonable to estimate the change of the bath von Neumann entropy as $\Delta S_B^{vN} \approx \beta \Delta \langle {\mathcal {\hat H}_B}\rangle +\beta( \Delta \langle {\hat {\mathcal H}_I} \rangle  - \delta U_A')$. Accordingly, we estimate the change of the total von Neumann entropy as $\Delta S_{tot}^{vN} \approx  \Sigma_{tot}^{vN}+\beta( \Delta \langle {\hat {\mathcal H}_I} \rangle  - \delta U_A')$, which leads to $\Delta I(A:B) \approx 0$.

\begin{figure}[tb]
    \centering
     \includegraphics[width = 1.0\textwidth]{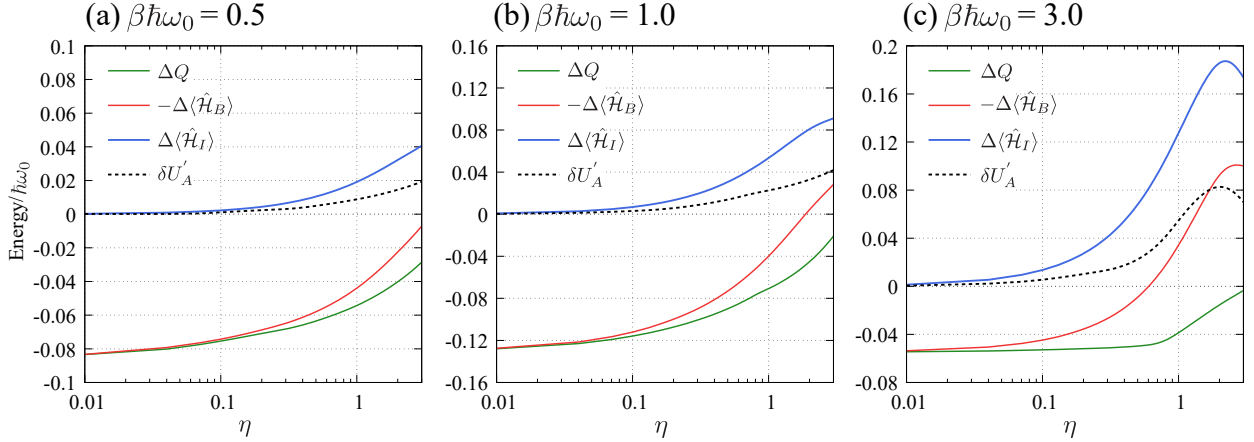}
     \vspace*{0.5cm}
     \caption{Change of the heat $\Delta Q$ (green curve), the decrease of the bath energy  $-\Delta \langle {\hat {\mathcal H}_B} \rangle$ (red curve), the change of the interaction energy $\Delta \langle {\hat {\mathcal H}_I} \rangle$ (blue curve), and the change of the system part of the system--bath interaction energy  $\delta U_A' \equiv \Delta U_A - \Delta \langle {\hat {\mathcal H}_A} \rangle$ (black dashed curve), plotted as functions of $\eta$ for (a) $\beta \hbar \omega_0 = 0.5$, (b) $\beta \hbar \omega_0 = 1$, and (c) $\beta \hbar \omega_0 = 3$.}
     \label{fig:Energy}
\end{figure}

To analyze this more closely, we depict $\Delta Q$, $-\Delta \langle {\hat {\mathcal H}_B} \rangle$, $\Delta \langle {\hat {\mathcal H}_I} \rangle$, and $\delta U_A' \equiv \Delta U_A - \Delta \langle {\hat {\mathcal H}_A} \rangle$ as functions of $\eta$ for various temperatures calculated from Eqs.~\eqref{eq:first},~\eqref{eq:bathE}, and~\eqref{eq:InteractionE} under the same physical conditions as in Fig.~\ref{fig:entropy}. As Figs.~\ref{fig:Energy}(a)--\ref{fig:Energy}(c) indicate, the total entropy production in the von Neumann case becomes negative due to the negative contribution of the bath entropy in the strong coupling region. In the low-temperature case, $\Delta Q$ and $-\Delta \langle {\hat {\mathcal H}_B}\rangle$ decrease after a maximum around $\eta=2$ because the strong dissipation suppresses heat transfer from the system to the bath.\cite{KATO2015} As a result, the entropy production in the von Neumann case slightly increases for $\eta>2$. 

In the Boltzmann case, $\Sigma_{tot}^{B}$ is always positive because the large positive contribution of the entropy from the system--bath interaction, $\beta( \Delta \langle {\hat {\mathcal H}_I} \rangle - \delta U_A')$,  compensates for the negative contribution of the bath entropy. This indicates that the total entropy production in the von Neumann case becomes negative because the contribution from the system--bath interaction has not been treated properly. Moreover, we find that $\delta U_A' \approx \Delta \langle {\hat {\mathcal H}_I} \rangle/2$, which indicates that the system--bath interaction energy is evenly distributed to the system and the bath. The difference of the bath entropy between the Boltzmann case and the von Neumann case is then evaluated as $\Delta Q - (-\Delta \langle {\hat {\mathcal H}_B}\rangle )\approx - \Delta \langle {\hat {\mathcal H}_I} \rangle/2$. Although the negativity of entropy production in the von Neumann case was found by Goyal, He, and Kawai,\cite{Kawai2019} here we identify the origin of this negativity using the Boltzmann entropy.

Finally, we discuss the characteristic features of the total entropy production in the Boltzmann case, depicted as the blue curves in Figs.~\ref{fig:entropy}(b-i)--\ref{fig:entropy}(b-iii). When $\eta$ is very weak, the system state ${\hat \rho_A}(t)$ at time $T$ is not quasi-static ${\hat \rho_A^{qeq}}(T)$ because $T=10/\omega_0$ is much shorter than the thermal relaxation time of the system in this parameter region, as with small $T$ in Fig.~\ref{fig:TD}. Thus, the total entropy production $\Sigma_{tot}^{B}$ becomes large for small $\eta$. The heat produced $\Delta Q=\Delta {U}_A-W$ reflects the nonequilibrium state of the system because, although the work $W$ defined by Eq.~\eqref{eq:work} was evaluated without assuming a quasi-static state, we evaluated $\Delta U_A$ and $\Delta S_A$ using ${\hat \rho_A^{qeq}}(t)$ through $\Delta F_A$ defined by Eq.~\eqref{eq:Fchange}. When the bath temperature becomes lower, the entropy production becomes larger, because thermal fluctuations, which help the relaxation to the equilibrium states, are suppressed.
As $\eta$ increases, $\Sigma_{tot}^{B}$ decreases because ${\hat \rho_A}(t)$ approaches ${\hat \rho_A^{qeq}}(t)$. If there is very strong system--bath coupling, the system and the bath are almost merged and behave like a single isolated system. Thus, we have $k_B^{-1}\Delta S_A = \beta \Delta Q$ and $\Sigma_{tot}^{B}$ becomes zero, which indicates that the dynamics of the total system is time reversible. 

In Appendix~\ref{sec:product}, we show that the total entropy production in the von Neumann case becomes positive, if we start from a factorized initial state. This is because the change of the bath entropy is enhanced, and it restores the loss of the entropy due to the factorized initial state. 

\subsection{Heat and the Jarzynski equality}
\label{sec:ZAproblem}

Although the equality $Z_{tot} (\tau)/Z_{tot}(0)=Z_A (\tau)/Z_A (0)$ is commonly assumed when investigating entropy production,\cite{Jarzynski2004,SeifertPRL16,Miller,Campisi09,Campisi11} it is obvious that this relation does not hold for an open quantum dynamics system in which fluctuations and dissipation play an essential role. This is because, although the first law of thermodynamics states that $\Delta Q(\tau)=\Delta {U}_A (\tau)-W(\tau)$, the above equality is equivalent to assuming that $\Delta {U}_A (\tau)-W(\tau)=0$ because
 \begin{align}
    \Delta {U}_A (\tau) \equiv -\frac{\partial}{\partial \beta} \ln \left( \frac{Z_A (\tau)}{Z_A (0)} \right),
\end{align}
and $W(\tau)= \Delta U_{tot}(\tau)$ with
\begin{align}
    \Delta U_{tot}(\tau) \equiv -\frac{\partial}{\partial \beta} \ln \left( \frac{Z_{tot} (\tau)}{Z_{tot} (0)} \right).
\end{align}
In the present case, we have 
$Z_{tot}(\tau)/Z_{tot}(0)=\left(Z_A (\tau) /Z_A (0)\right)\left(Z_B(\tau)/Z_B(0)\right)$ and $Z_{tot} (\tau)/Z_{tot}(0) \ne Z_A (\tau)/Z_A (0)$, where $\left( Z_B(\tau)/Z_B(0)\right)$ is evaluated from $\Delta Q(\tau) = {\partial}\ln \left( {Z_{B} (\tau)}/{Z_{B} (0)}\right)/{\partial \beta} $.
As a result, the Jarzynski equality does not hold.  This is natural, because the situation we consider here is not adiabatic (i.e., $\Delta Q(\tau) \ne0$) and not time reversible, although the total energy of the system plus bath is still conserved, as described by Eq.~\eqref{eq:Work_Htot}.

\section{CONCLUDING REMARKS}
 \label{sec:conclusion}

In this paper, we present a quantitative scheme to evaluate thermodynamic variables, such as the change of the Boltzmann entropy, for isothermal processes in an open quantum dynamics system. The scheme is based on evaluating the quasi-static Helmholtz energy using a reduced equation of motion for any system coupled to a heat bath under a slowly changing external force. Any open quantum dynamics formalism that can accurately describe the thermal equilibrium state as a steady-state solution can be employed for the calculations. Because the present approach is based on the kinetic response of a thermal system, it may be possible to apply a similar scheme in a real experiment with a small quantum system in a thermal environment by applying a time-dependent external perturbation. 

As a demonstration, we calculated various thermodynamics valuables for a spin-boson system. We find that, although the profiles of the system entropy in the Boltzmann and von Neumann cases as functions of the system--bath coupling strength are similar, those for the total entropy production are completely different. The total entropy production in the Boltzmann case is always positive, whereas that in the von Neumann case becomes negative if we chose a thermal equilibrium state of the full system (a correlated thermal state) as the initial condition. This is because the total entropy production in the von Neumann case does not properly take into account the contribution of the entropy from the system--bath interaction. Finally, the applicability of the Jarzynski equality is briefly discussed based on partition functions.

Although the differences between the results for the Boltzmann case and the von Neumann case are small in a region with weak system--bath coupling, a formalism based on the Boltzmann entropy must be used to investigate the philosophical foundations of quantum thermodynamics, and there should be no inconsistencies. Moreover, ignoring the effect of the system--bath interaction is unrealistic because a tiny quantum system can never reach thermal equilibrium on its own without the system--bath interaction.

In the present paper, although we limited our analysis to a simple spin-boson system, we can use the same approach for the variety of systems that the HEOM formalism has been applied to.\cite{YTperspective} Moreover, if we employ the quantum hierarchical Fokker--Planck equations (QHFPEs) for a system described by a configuration space and Wigner distribution functions,\cite{TanimuraPRA91,TanimuraJCP92,TanimuraJPSJ06,YTJCP2015,KatoJPCB13} we can investigate not only quantum cases but also classical cases by taking the classical limit of the QHFPEs. Because the QHFPE formalism treats quantum and classical systems in the same way, regardless of the form of the potential, it can be used to identify purely quantum mechanical effects by comparing the classical and quantum results for the Wigner distribution.\cite{TanimuraJCP92,KatoJPCB13} In conclusion, the present paper provides a rigorous and quantitative framework for investigating quantum thermodynamics.

\begin{acknowledgments}
    The financial support received from the Kyoto University Foundation is gratefully acknowledged.
\end{acknowledgments}
    
\section*{Data availability}
The data that support the findings of this study are available from the corresponding author upon reasonable request.

\appendix
\section{Derivation of Eq.~(\ref{eq:qspower})}
\label{sec:qspower}

In this appendix, we derive Eq.~(\ref{eq:qspower}) for the system--bath Hamiltonian expressed as ${\mathcal {\hat H}}(\tau) = {\mathcal {\hat H}}_{0} + {\mathcal {\hat H}}_E (\tau)$, where ${\mathcal {\hat H}}_0\equiv{\mathcal {\hat H}}_A^0 + {\mathcal {\hat H}}_I + {\mathcal {\hat H}}_B$ and ${\mathcal {\hat H}}_E (\tau)$ is the time-dependent part of the system Hamiltonian.  
Using Kubo's identity,\cite{KuboBook} we can rewrite the partition function of the Hamiltonian  ${\mathcal {\hat H}}(\tau + \Delta \tau)$, expressed in real time, as:
\begin{align}
    e^{-\beta {\mathcal {\hat H}}(\tau + \Delta \tau)} = e^{-\beta {\mathcal {\hat H}}(\tau)} \left[ 1 - \int_0^{\beta} d\lambda e^{\lambda {\mathcal {\hat H}}(\tau)} \Delta {\mathcal {\hat H}_E}(\tau) e^{-\lambda ({\mathcal {\hat H}}(\tau) + \Delta {\mathcal {\hat H}_E}(\tau))} \right],
    \label{eq:KuboIdentity}
\end{align} 
where $\Delta {\mathcal {\hat H}_E}(\tau) \equiv {\mathcal {\hat H}_E}(\tau + \Delta \tau) - {\mathcal {\hat H}_E}(\tau)$. For small $\Delta {\mathcal {\hat H}_E}(\tau)$, which is realized when ${\mathcal {\hat H}_E}(\tau)$ changes in time slowly or $\Delta \tau$ is small, we can ignore the higher-order contribution of $\Delta {\mathcal {\hat H}_E}(\tau)$. For the reduced density operator in imaginary time, defined as ${\hat {\tilde \rho}}_A (\beta \hbar; \tau) \equiv {\operatorname{tr}_B} \{ e^{-\beta {\mathcal {\hat H}}(\tau)} / Z_B^{\tau} \}$,
\cite{YTJCP2014,YTJCP2015} we then have:
\begin{align}
    {\hat {\tilde \rho}}_A (\beta \hbar; \tau + \Delta \tau) &\approx {\hat {\tilde \rho}}_A (\beta \hbar; \tau) - \int_0^{\beta} d\lambda \operatorname{tr}_B \left\{ \frac1{Z_B^{\tau}} e^{-\beta {\mathcal {\hat H}}(\tau)} e^{\lambda {\mathcal {\hat H}}(\tau)} \Delta {\mathcal {\hat H}_E}(\tau) e^{-\lambda {\mathcal {\hat H}}(\tau)} \right\}.
    \label{eq:DifferentialImDensity}
\end{align}
The reduced partition function of the system is given by $Z_A (\tau) = \operatorname{tr}_A \{ {\hat {\tilde \rho}}_A (\beta \hbar; \tau) \}$. Thus, for a slowly changing time-dependent Hamiltonian with $\Delta \tau \rightarrow 0$, we have the relation:
\begin{align}
    \frac{\partial}{\partial \tau} Z_A (\tau) = -\int_0^{\beta} d\lambda \operatorname{tr}_A \left[ {\operatorname{tr}_B} \left\{ e^{-\beta {\mathcal {\hat H}}(\tau)} \frac1{Z_B^{\tau}} e^{\lambda {\mathcal {\hat H}}(\tau)} \frac{\partial {\mathcal {\hat H}_E}(\tau)}{\partial \tau} e^{-\lambda {\mathcal {\hat H}}(\tau)} \right\} \right].
      \label{eq:ImDensityDerivative}
\end{align}
By dividing both sides of the equation by $Z_A (\tau)$, we can write Eq.~(\ref{eq:qspower}) as:
\begin{align}
    \frac{\partial}{\partial \tau} \left( -\frac1\beta \ln Z_A (\tau) \right) &=\frac1\beta \int_0^{\beta} d\lambda \operatorname{tr}_{A+B} \left\{ \frac1{Z_A (\tau) Z_B^{\tau}} e^{-\beta {\mathcal {\hat H}}(\tau)} e^{\lambda {\mathcal {\hat H}}(\tau)} \frac{\partial {\mathcal {\hat H}_E}(\tau)}{\partial \tau} e^{-\lambda {\mathcal {\hat H}}(\tau)} \right\} \nonumber\\
    &= {\operatorname{tr}_{A+B}} \left[ \frac1{Z_A (\tau) Z_B^{\tau}} e^{-\beta {\mathcal {\hat H}}(\tau)} \frac{\partial {\mathcal {\hat H}_E}(\tau)}{\partial \tau} \right] \nonumber\\
    &= {\operatorname{tr}_A} \left\{ {\hat \rho}_A^{qeq} (\tau)  \frac{\partial {\mathcal {\hat H}_E}(\tau)}{\partial \tau} \right\},
    \label{eq:ZADerivative}
\end{align}
where the quasi-thermal equilibrium state of the system at time $\tau$ is defined as ${\hat \rho}_A^{qeq} (\tau) \equiv {\hat {\tilde \rho}}_A (\beta \hbar; \tau)/Z_A (\tau).$\cite{YTJCP2014,YTJCP2015}

\section{Total entropy production from the factorized initial state}
\label{sec:product}

To illustrate the contribution of the entropy from the system--bath interaction, here we present the results in the von Neumann case calculated from the factorized thermal initial state, ${\hat \rho}_{tot} (0) = {\hat \rho}_{A}^{(eq)} \otimes {\hat \rho}_{B}^{eq}$, using Eq.~\eqref{eq:Neumann-entropyt}. Note that, although the factorized initial state has been intensively used to investigate entropy production, it is valid only for a Markovian heat bath, as the noise correction of the bath is short ($\gamma \gg\omega_0$) and the temperature is very high ($\beta\hbar\omega \ll 1$) or the system--bath interaction is very weak ($\eta \ll \omega_0$).\cite{TanimuraJPSJ06,YTJCP2014,YTJCP2015} At low temperatures, where quantum effects play a dominant role, non-Markovian effects arise, even for $\gamma \gg\omega_0$, due to the quantum thermal fluctuations, as observed in the simulation of muon spin spectroscopy ($\mu$SR).\cite{TakahashiJPSJ20}

\begin{figure}[tb]
    \centering
     \includegraphics[width = 0.8\textwidth]{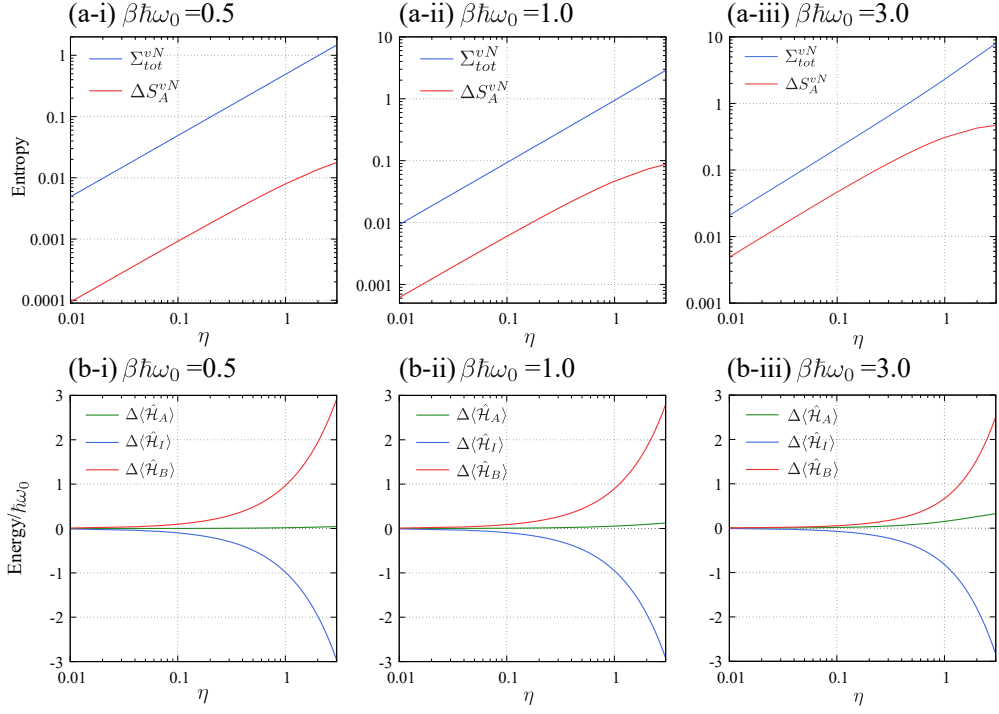}
     \caption{(a) Total entropy production (blue curve) and the change of the system entropy (red curve). (b) Change of the system energy (green curve), interaction energy (blue curve), and bath energy (red curve). These were calculated from the factorized initial condition in the von Neumann case as functions of the system--bath coupling for (i) $\beta \hbar \omega_0 = 0.5$, (ii) $\beta \hbar \omega_0 = 1.0$, and (iii) $\beta \hbar \omega_0 = 3.0$, under the same physical conditions as in Figs.~\ref{fig:entropy} and~\ref{fig:Energy}.}
     \label{fig:product}
\end{figure}

In Fig.~\ref{fig:product} we present the results of (a) the change of the system entropy and total entropy production, and (b) the change of the system energy, interaction energy, and bath energy in the von Neumann case calculated from the factorized initial state under the same physical conditions as in Figs.~\ref{fig:entropy} and~\ref{fig:Energy}. Unlike the correlated case in Fig.~\ref{fig:entropy}(b), $\Sigma_{tot}^{vN}$, as shown  in Fig.~\ref{fig:product}(a), is always positive for any strength of the system--bath coupling. Although this result is consistent with former investigations,\cite{Sagawa12, EspositoNJP10} it is due to the contribution of the entropy from the system--bath interaction, which has not been considered before. Although the system energy $\Delta \langle {\hat {\mathcal H}_A} \rangle$ does not change significantly, regardless of $\eta$, the change of the interaction energy $\Delta \langle {\hat {\mathcal H}_I}\rangle $ dramatically decreases to restore the system--bath correlation that is lost from the factorized initial state, as illustrated in Fig.~\ref{fig:product}(b). Then the bath energy $\Delta \langle {\hat {\mathcal H}_B} \rangle$, which contributes to $\Sigma_{tot}^{vN}$ through the second term on the RHS of Eq.~\eqref{eq:Neumann-production}, increases and supplies energy to the system--bath interaction. 

As explained in Sec.~\ref{subsec:EntropyProduction}, the total entropy production in the von Neumann case is 
underestimated by about $\Delta \langle {\hat {\mathcal H}_I} \rangle/2$. In the present factorized case, 
because $\Delta \langle {\hat {\mathcal H}_I} \rangle$ is negative, the modified total entropy production 
${{\Sigma}_{tot}^{vN}}'= \Sigma_{tot}^{vN} + \beta \Delta \langle {\hat {\mathcal H}_I} \rangle/2$ becomes smaller than the original value, 
whereas ${{\Sigma}_{tot}^{vN}}'$ is still positive.

\bigskip

\renewcommand{\section}[2]{}

\end{document}